\documentclass[jcp,preprint,superscriptaddress]{revtex4}

\usepackage[dvips]{graphicx}
\usepackage{amsmath}
\usepackage{amsfonts}
\usepackage{amssymb}
\usepackage{dcolumn}
\usepackage{bm}
\usepackage{aeguill}
\usepackage[ansinew]{inputenc}
\usepackage{latexsym}
\usepackage{color}
\usepackage{epsfig}
\begin{document}

\title{Scanning tunneling microscopy simulations of poly(3-dodecylthiophene) chains adsorbed on highly oriented pyrolytic graphite}

\date{\today}

\author{M. Dubois}
\email{mathieu-j.dubois@cea.fr}
\affiliation{CEA/DRFMC/SPSMS/GT, 17 rue des Martyrs, 38054 Grenoble Cedex 9, France}
\author{S. Latil}
\affiliation{Laboratoire de Physique du Solide, FUNDP, 61 rue de Bruxelles, B-5000 Namur, Belgium}
\author{L. Scifo}
\author{B. Grévin}
\affiliation{UMR5819 (CEA-CNRS-Université Grenoble I), CEA/DRFMC/SPrAM/LEMOH, 17 rue des Martyrs, 38054 Grenoble Cedex 9, France}
\author{Angel Rubio}
\altaffiliation[Permanent address: ]{Departamento de F\'{\i}sica de 
Materiales, Facultad de Qu\'\i micas Universidad del Pa\'\i s Vasco, 
Centro Mixto CSIC-UPV, and Donostia International Physics Center (DIPC), 
E-20018 Donostia-San Sebasti\'an, Spain}
\affiliation{Institut f\"ur Theoretische Physik, Freie Universit\"at Berlin, Arnimallee 14, D-14195 Berlin, Germany}
\affiliation{European Theoretical Spectroscopy Facility (ETSF)}

\begin{abstract}

We report on a novel scheme to perform efficient simulations of Scanning Tunneling Microscopy (STM) of molecules weakly bonded to surfaces. Calculations are based on a tight binding (TB) technique including self-consistency for the molecule to predict STM imaging and spectroscopy. To palliate the lack of self-consistency in the tunneling current calculation, we performed \textit{first principles} density-functional calculations to extract the geometrical and electronic properties of the system. In this way, we can include, in the TB scheme, the effects of structural relaxation upon adsorption on the electronic structure of the molecule. This approach is applied to the study of regioregular poly(3-dodecylthiophene) (P3DDT) polymer chains adsorbed on highly oriented pyrolytic graphite (HOPG). Results of spectroscopic calculations are discussed and compared with recently obtained experimental data.

\end{abstract}

\maketitle

\section{Introduction}

The Scanning Tunneling Microscope is a remarkable tool to probe objects and materials at the atomic scale and has then widely been used to characterize the adsorption of various molecules on metallic or semiconducting surfaces\cite{Chavy93,Lopinski98,Yokoyama01,Lastapis05,Lu04}. Moreover, STM can be used to understand the relations between the electronic properties of materials and their structure at atomic scale. Thus, the analysis of STM images allows, i) to study the structural parameters of the system; ii) to study the nature of the probed electronic states through the analysis  of 
the bias dependent STM images. Also, complementary, Scanning Tunneling Spectroscopy (STS) measurements allow to scan the local density of states of the system\cite{Tersoff83,Stroscio86}. However, the correspondence between topographic or spectroscopic measurements, and the underlying electronic structure is not always obvious. The bonding nature between the molecule and the substrate, the charge transfer or the interaction between severals molecules are examples of parameters that can modify the electronic properties of the system with respect to its isolated parts. There is therefore a crucial need for efficient theoretical tools to tackle these issues. Most of the models developed so far are based either on a simplified perturbative Bardeen approach\cite{Hofer01} or on the Landauer formalism\cite{Ness91}. These methods have been successfully developed on the basis of \textit{ab initio} or semi-empirical techniques on individual molecules\cite{Hofer01,Sautet91} or small supramolecular systems\cite{Toerker02} but the main restriction, at the present time, remains the computing time needed for those calculations. The question of the suitability of such methods for describing large molecular systems can then be debated.\\

In this paper, an alternative model combining both \textit{ab initio} and semi-empirical approaches for the simulation of STM experiments is presented. This method is particularly well suited for the case of physisorbed molecules for which the effects of the adsorption on the electronic structure of the molecule remain weak. Thus, the tunneling current can be calculated within the tight binding approximation with no need for self-consistency. However, information about the geometrical structure and the electronic properties of the system remains essential and will first be extracted from first principles calculations. This method will be discussed and compared with STS spectra recently obtained for regioregular poly(3-dodecylthiophene) (P3DDT) chains adsorbed on highly oriented pyrolytic graphite (HOPG). Indeed, $\pi$-conjugated semi-conducting polymers have emerged as a new class of materials which present both self-assembly and novel electronic features\cite{Okawa01,Akai03}. The physics of poly(3-alkylthiophenes), which stand as a generic model owing to their remarkable semi-crystalline properties and resulting high carrier mobilities\cite{Sirringhaus99} can then directly be addressed by STM. Therefore, a recent study of regioregular P3DDT adsorbed on HOPG has been devoted to determine, at the local scale, the precise relationship between structural organization and electronic properties. Details about these experiments will be published elsewhere. The aim of this paper is not to give a complete theoretical analysis of this experimental study but to present a generic method that can be applied to various systems. We will then focus on an application of our theoretical model to the study of STS spectra on single defectless P3DDT chains adsorbed on a graphene surface.

\section{Description of the methods}

\subsection{Tight Binding Calculation of the Tunneling Current}

STM experiments are simulated by calculating the current between a tip, that is assumed to have a pyramidal apex\cite{Perdigao04}, and a graphite surface, on which is physisorbed a single chain of P3DDT, as a function of the applied bias $V_g$ to the STM tip. The calculation is based on a tight binding formulation of the elastic scattering theory considering that the Hamiltonian of the system can be written as $H=H_0+V$ where $H_0$ is the Hamiltonian of the three uncoupled regions (the STM tip, the molecule, and the substrate) and $V$ their coupling\cite{DelerueBook}. The current is then given by

\begin{eqnarray}	
\nonumber I(V_g)=\frac{2\pi e}{\hbar} \sum_{i\in S,j\in T}\left|\left\langle j \right|T(\epsilon_i)\left|i\right\rangle \right|^2 \\
\times \left\{f(\epsilon_j-\mu_T)-f(\epsilon_i-\mu_S)\right\}\delta(\epsilon_i-\epsilon_j),
\label{eq_current}
\end{eqnarray}

\noindent where $\left|j\right\rangle$ and $\left|i\right\rangle$ are the eigenstates of $H_0$ in the tip (T) and in the substrate (S), respectively, $\epsilon_j$ and $\epsilon_i$ being their energies. Equation (\ref{eq_current}) assumes that the eigenstates of $H$ incident from the surface and partially transmitted to the tip are occupied by electrons up to the Fermi level $\mu_S$, and symmetrically those incident from the tip are filled up to $\mu_T$ ($f$ is the Fermi-Dirac distribution function and $\mu_T-\mu_S=eV_g$). As shown in Refs.\onlinecite{DelerueBook} and \onlinecite{Todorov02}, Eq.(\ref{eq_current}) is a generalization of the Landauer formula. The scattering operator $T(\epsilon)$ is expressed at each energy $\epsilon$ in a Green's function formalism\cite{Datta97} as 

\begin{equation*}
T(\epsilon)=V+VG(\epsilon)V,
\end{equation*}

\noindent where $G$ is the Green's function of the coupled system obtained from the Green's function of the isolated regions 

\begin{equation*}
G_0(\epsilon)=\lim_{\gamma \to 0}(\epsilon-H_0+i\gamma)
\end{equation*}

\noindent

via the Dyson's equation

\begin{equation*}
	G=[I-G_0V]^{-1}G_0.
\end{equation*}

\noindent A small imaginary part $\gamma$ is added in the Green's functions to avoid the divergence for energies corresponding to the poles of the Hamiltonian. It also have a physical meaning, and reflects the broadening of the electronic states due to the coupling between the different parts. Considering that the interaction of the molecule with the two electrodes is small, we take $\gamma=0.05$ eV.

For the coupling of the molecule with the tip and the surface, the hopping terms $V_{\alpha\beta}$ between two atomic orbitals $\left|\alpha\right\rangle$ and $\left|\beta\right\rangle$ follow an exponential decay with the inter-atomic distance $d$. The prescription of Ref.\onlinecite{Lefebvre98} is assumed so that

\begin{equation}
	V_{\alpha\beta}(d)=\exp\left[-2.5\left(\frac{d}{d_0}-1\right)\right]V_{\alpha\beta}^H(d_0),
\end{equation}

\noindent where $d_0$ is the sum of the covalent radii of the two atoms and $V^H_{\alpha\beta}(d_0)$ is the hopping integral calculated from Harrison's rules\cite{Harrison} at distance $d_0$.\\

To compute the Green's functions needed to calculate the tunneling current, the electronic structure of the system has first to be determined. Considering that the molecule is only weakly interacting with the surface and the tip, as it is the case in physisorption, one can reasonably assume that its electronic structure will not be strongly affected by adsorption. On the other hand, electronic correlations inside the molecule are important and have to be taken into account. Hence, the electronic structure of the molecule is calculated using a self-consistent TB method described by Krzeminski et \textit{al}\cite{Krzeminski99,Krzeminski01} and electron-electron interactions are treated by a method derived from the orthodox theory\cite{DelerueBook}. In order to take into account charging effects (i.e., the electrons injected by the electrodes are charging the molecule), the molecule has to be considered as an electronic quantum dot\cite{Niquet02}, characterized by affinity and ionization levels, $\epsilon_i^e(V_g)$ and $\epsilon_j^h(V_g)$, that amount to the injection of one electron or one hole, respectively. The relevant gap for STM simulations is then given by the quasi-particle gap (i.e., the gap obtained from a many body approach and that is different to the Kohn-Sham eigenvalues) defined as the difference between Electron Affinity (EA) and Ionization Potential (IP). Moreover, the injection of an extra charge inside the molecule will also lead to a response of the charges inside the metallic electrodes. This extra charge will then be screened by the graphite surface and the STM tip. This phenomenon, that tends to reduce the gap of the molecule, is taken into account by the image charge method\cite{Krzeminski01-2}. Finally, the dependence of the molecular levels with the applied bias is described by

\begin{equation}
	\epsilon_i^{e(h)}(V_g)=\epsilon_i^{e(h)}(V_g=0)-\eta eV_g
	\label{pot_drop}
\end{equation}

\noindent where $\eta$ is the same constant for all the electronic states\cite{Krzeminski01-2}. The parameter $\eta$ is determined by the geometry of the junction and describes the potential's drop between the two electrodes. $\eta$ can take values between 0 and 1, the 0.5 value corresponding to a symmetric interaction of the molecule with the two leads, and the lower and upper limits standing for a molecule whose coupling to one electrode is predominant\cite{Datta05}. This parameter (which is the only tunable parameter of our model), as well as the equilibrium energetic state of the junction ($V_g=0$) will determine the electronic states involved in the tunneling process.

\subsection{\textit{Ab initio} calculations}

Although the previously described TB scheme is sufficient to perform STM simulations, \textit{ab initio} calculations can provide important information about the structural and electronic properties of the system. Indeed, we will use first principles calculations to extract the real adsorption configuration of the molecule on the surface and to determine the effects of the adsorption on the electronic structure of the molecule. We will show how these information can directly be integrated into the TB model.\\

\begin{figure}[h!]
\includegraphics[width = 0.35\columnwidth]{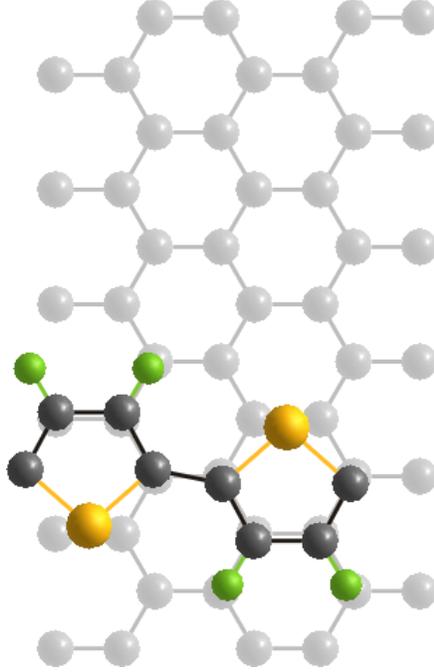}
\caption{(Color online) Relaxed structure of the periodic system (polymer adsorbed on a graphene layer) obtained by imposing that the dodecyl chains of the polymer (not shown) match the zigzag pattern of the graphite lattice. In the perpendicular direction, the periodicity of the surface is taken large enough to avoid interactions between the $\pi$-conjugated part of molecules in neighboring cells. Carbon, sulphur and hydrogen atoms are represented using gray, yellow and green colors, respectively.}
\label{molecule}
\end{figure}

The structural and electronic properties of the adsorbed polymer on the graphite surface have been calculated by \textit{ab initio} density-functional theory (DFT) simulations using a plane wave basis set treating the electron-ion interaction within the projector augmented wave method (PAW) and the Generalized Gradient Approximation (GGA) for describing exchange-correlation effects as implemented in the VASP code.\cite{VASP} The relaxed structure has first been obtained by energy minimization of the periodic unit of the polymer chain (corresponding to two thiophene rings) on a graphene layer. The molecule has been positioned at 0.35 nm from the surface, which corresponds to typical distance for $\pi$-$\pi$ interactions. These calculations were performed taking into account that the dodecyl chains of the polymer are epitaxed onto the zigzag pattern of the honey comb lattice of HOPG, as already experimentally evidenced.\cite{Mena00,Azumi00} This imposes that the periodicity of the molecule matches the one of the graphene layer, and we will then consider the periodic unit of the system represented on Fig.\ref{molecule} for the rest of the \textit{ab initio} calculations. Thus, the considered unit cell comprises a large enough graphene layer to avoid interaction between molecules of adjacent cells in the perpendicular direction of the polymer chain. Finally, to reduce the computional cost, the periodic unit of P3DDT will be considered without the dodecyl chains as they do not contribute to the tunneling current. Indeed, alkyl chains, made of $\sigma$ bonds between carbon atoms, are very insulating as the HOMO of the chain is several eV below the Fermi level of the surface. Thus, the dodecyl chains' properties will not been probed for voltages generally applied in STM experiments.\\

\begin{figure*}[ht!]
\includegraphics[width = 2.0\columnwidth]{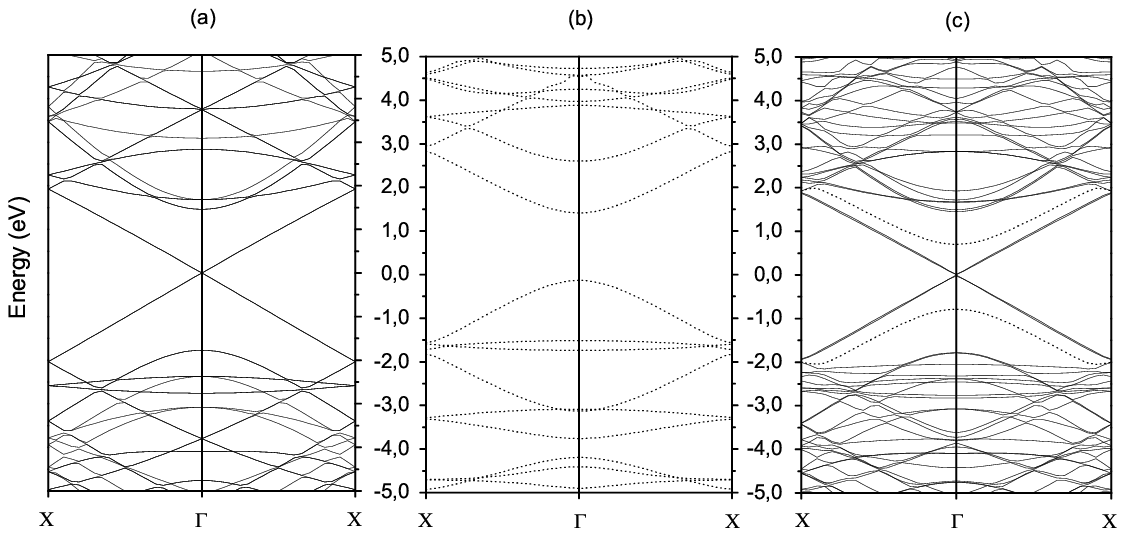}
\caption{Calculated band structures for (a) the graphene surface, (b) the isolated polymer and (c) the complete polymer and graphite system. On the latter, the HOMO and LUMO states of the polymer are identified by dotted lines. The zero of energy in each panel is settled by the respective Fermi level of each system. The adsorption of the polymer on the surface only leads to a rigid shift of all molecular states.}
\label{band_struc}
\end{figure*}

We now need to extract several parameters, for the TB model, as far as the electronic structure of the molecule is concerned. We start by determining what are the effects of the adsorption on the electronic structure of the molecule. As already mentioned, in the case of a weak bonding with the surface, these effects should be small. In order to check this point, we look to dispersionless features in the band structure of the combined system.
Indeed, since the system is periodic in one direction of space, it is possible to extract the band structure of both isolated surface and molecule and compare them to the one of the complete system. The three band structures, plotted along the $\Gamma$-$X$ direction in the first Brillouin zone, are represented in Fig.\ref{band_struc}. The results confirm that the adsorption of P3DDT on HOPG is relatively weak and that it does not modify the electronic structure of the polymer except for a shift of all the molecular states so that there is almost an alignment of the midgap of the polymer with the Fermi level of the graphite surface.\\
This also suggests that no significant charge transfer effects from the surface to the polymer layer should be expected. Finally, this justifies that the electronic structure of the three regions of the tunneling junction can be calculated separately for the calculation of the current. One just has to rigidly shift all the molecular levels to keep the same relative position of occupied and unoccupied states of the polymer with respect to the Fermi level of the surface (with a work function of graphene of $\phi$ = 4.5 eV below vacuum level).\\

\begin{figure}[h!]
\includegraphics[width = 1.0\columnwidth]{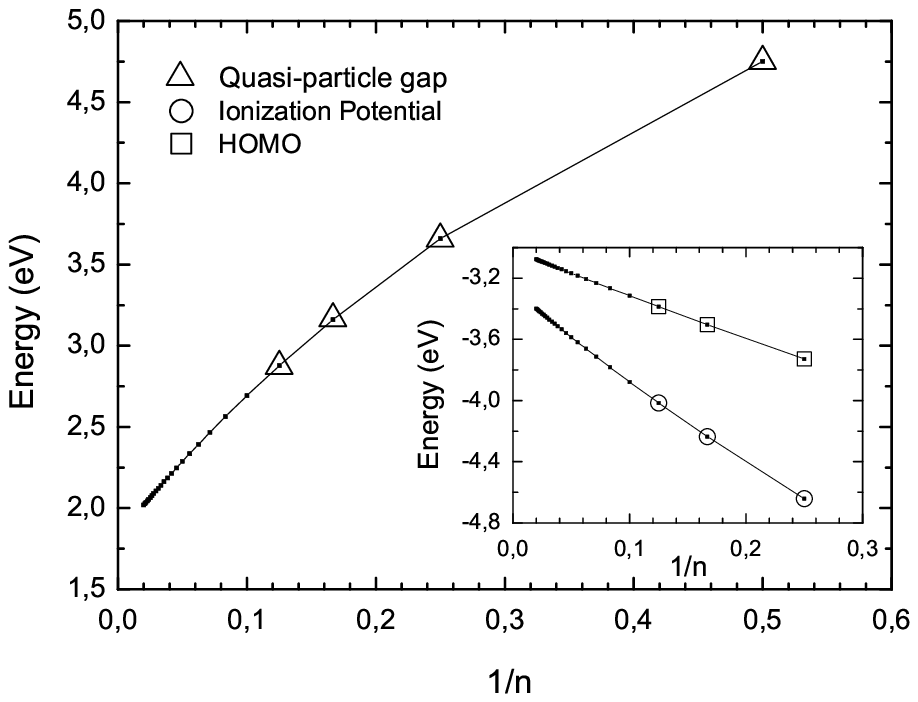}
\caption{Evolution of the calculated quasi-particles gap (EA-IP) of poly(thiophene)s as a function of the inverse of the number of rings ($\triangle$) and the corresponding fitting curve. Inset: evolution of HOMO ($\square$) and IP ($\circ$) levels as a function of the inverse of the number of rings and their fitting curve. Calculated values have been fitted by a function $a_1+{a_2\over{n+a_3}}$, where $a_1,a_2,a_3$ are fitting parameters and $n$ the number of thiophene rings. }
\label{quasi-part}
\end{figure}

However, in the TB method developed to calculate the tunneling current, a finite sized system is required. Since the electronic gap of $\pi$-conjugated polymers decreases and, at some point, saturates when increasing the number of periodic units\cite{Telesca01}, it is possible to determine what are the number of thiophene rings needed to replace an infinite long polymer chain by a finite one with similar quasi-particle gap. This can be done by calculating the evolution of the gap of the polymer with the number of repeatition units. Hence, for a set a finite sized molecules (2,4,6 and 8 periodic units), the quasi-particle gap is calculated as the difference between Electron Affinity (EA) and Ionization Potential (IP) defined as

\begin{eqnarray}
\nonumber	EA=E[-1]-E[0]\\
IP=E[0]-E[+1]
\end{eqnarray}

\noindent where $E[-1]$, $E[0]$ and $E[+1]$ are the total energies of the negatively charged, neutral and positively charged species, respectively. The evolution of the quasi-particle gap as a function of the inverse of the number of repeating unit is presented in Fig.\ref{quasi-part}. The results obtained for the four first polythiophenes have then been fitted to predict the evolution for a large number of rings. Fig.\ref{quasi-part} shows that the quasi-particle gap of these compounds saturates for a large number of thiophene groups. One finds that above 18 rings (octadecathienyl) the value of the gap does not evolve significantly. This molecule will therefore be taken as a good approximation to simulate the infinite polymer chain. The value of the HOMO-LUMO gap (at the Kohn-Sham level of DFT) of the infinite polymer is $1.49$ eV (see Fig.\ref{band_struc} b), that should be compared to the molecular gap ($EA-IP=2.33$ eV for octadecathienyl) in order to get a glimpse of charging effects in the molecular system. In practice, taking into account the size of the considered octadecathienyl molecule, the graphite surface will be simulated by a single graphene sheet using 150 $k$ points in a super-cell made of 504 atoms and only one $\pi$ orbital by carbon atom will be kept ($V_{pp\pi}=-2.75$ eV, for the TB parametrization).

\begin{figure}[h!]
\includegraphics[width = 0.9\columnwidth]{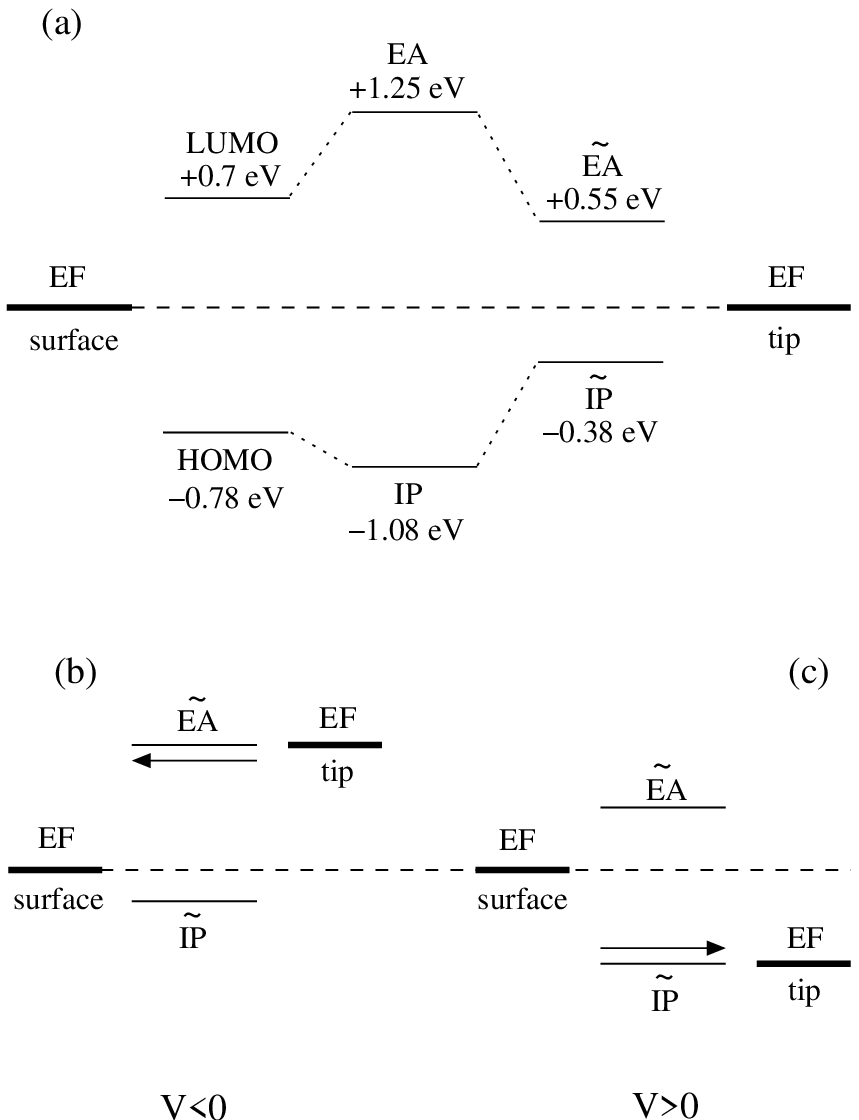}
\caption{Energetic schemes of the tunneling junction. (a) Energy levels of octadecathienyl molecule with respect to the Fermi levels of the electrodes at zero bias : single-particle levels for the HOMO and LUMO, ionization levels corresponding to the charging of the molecule with one electron (EA) or one hole (IP) for the isolated molecule and for the molecule coupled to the electrodes ( $\tilde{EA}$ and $\tilde{IP}$, respectively). (b), (c) Evolution of the bias-dependent molecular level positions taking $\eta=0.33$ (see text) : for a high enough negative bias, electrons are injected from the tip to the surface via $\tilde{EA}$ level of the molecule, whereas for a sufficiently positive bias electrons are injected form the surface to the tip via the $\tilde{IP}$ level of the molecule. }
\label{elec_levels}
\end{figure}

One should keep in mind that the band structures calculations for the extended system can only been performed for a neutral configuration. This means that we are only able, at this stage, to locate the HOMO (occupied states) and the LUMO (unoccupied states) of the molecule with respect to the Fermi level of the graphene surface, mimicking in a situation where no current is passing through the molecule. Notwithstanding, as already mentioned, the relevant gap for STM measurements is the so-called quasi-particles gap. Therefore, for those STM simulations the key quantity is to position all the molecular levels with respect to the Electron Affinity and Ionization Potential of the molecule. This study can be done easily by considering isolated polythiophenes and look at the evolution of the position of the HOMO and IP levels (with respect to the vacuum level) with the number of thiophene rings. The inset of Fig.\ref{quasi-part} shows that not only charging effects tend to reduce the molecular gap but that they also shift downwards all occupied states by a value of $\Delta E=-0.3$ eV. Assuming that the same change in the IP occurs when the polymer is adsorbed on the surface, then all occupied states have to be shifted by the same value $\Delta E$. A scissor operator is then applied to all unoccupied states so that the energy difference between occupied and unoccupied ones corresponds to the quasi-particle gap which value is $2.33$ eV for the octadecathienyl molecule. All this energy level scheme is schematically illustrated in Fig.\ref{elec_levels}a.\\
The subsequent relevant effect for the modeling of the STM current is the screening of the injected charge by the metallic surface and STM tip. This phenomenon, that tends to reduce the molecular gap, is taken into account by the image charge method. The renormalised last occupied molecular level and the first unoccupied one are now referred to as $\tilde{IP}$ and $\tilde{EA}$. The new relevant energetic situation is depicted in Fig.\ref{elec_levels}a.\\

From a practical point of view, in order to incorporate all those effects in the TB calculation of the tunneling current, we have to follow five steps :(i) to determine the electronic structure of the octadecathienyl molecule with the self-consistent TB method described in Ref.\onlinecite{Krzeminski99} and \onlinecite{Krzeminski01}; (ii) to rigidly shift all the molecular states so that the position of the midgap of the molecule with respect to the Fermi level of the graphite surface corresponds to the situation extracted from the band structure plotted on Fig.\ref{band_struc}c; (iii) to apply a shift of $\Delta E=-0.3$ to all occupied states so that the last occupied one corresponds to the IP of the molecule; (iv) to apply a scissor operator to all unoccupied states so that the molecular gap corresponds to the quasi-particle gap (the position of the first unoccupied state now corresponds to the EA of the molecule); (v) finally, to equally reduce the gap by the value extracted from the image charge method described in Ref.\onlinecite{Krzeminski01-2} (in the present case, this corresponds to a reduction of the levels of $0.7$ eV).   

\subsection{Experiments}

P3DDT films with sub-monolayer coverage were prepared as described in Ref.\onlinecite{Brun04}. STM experiments were performed under ultrahigh-vacuum conditions (base pressure below $5.10^{-11}$ mbars) using a VT Omicron system and mechanically prepared PtIr tips. Measurements were recorded at room temperature in the low-current mode (LC-STM) using an adequate preamplifier which operation required bias voltage (Vg) to be applied to the tip and sample to be grounded. Spectroscopic data were acquired in the current imaging tunneling spectroscopy mode (CITS) in which topographic images, in the constant current mode, are recorded simultaneously with $I(V_g)$ curves taken on a $100 \times 100$ points grid, feedback loop being disabled during spectroscopic acquisition. Bias voltage was swept starting from the topographic regulation set point. Conductance spectra $dI/dV_g$ were recorded using a lock-in technique, with a topographic feedback loop gain adjusted to avoid oscillations in the z-images.

\section{Results and discussion}

\begin{figure}[ht!]
\includegraphics[width = 1.0\columnwidth]{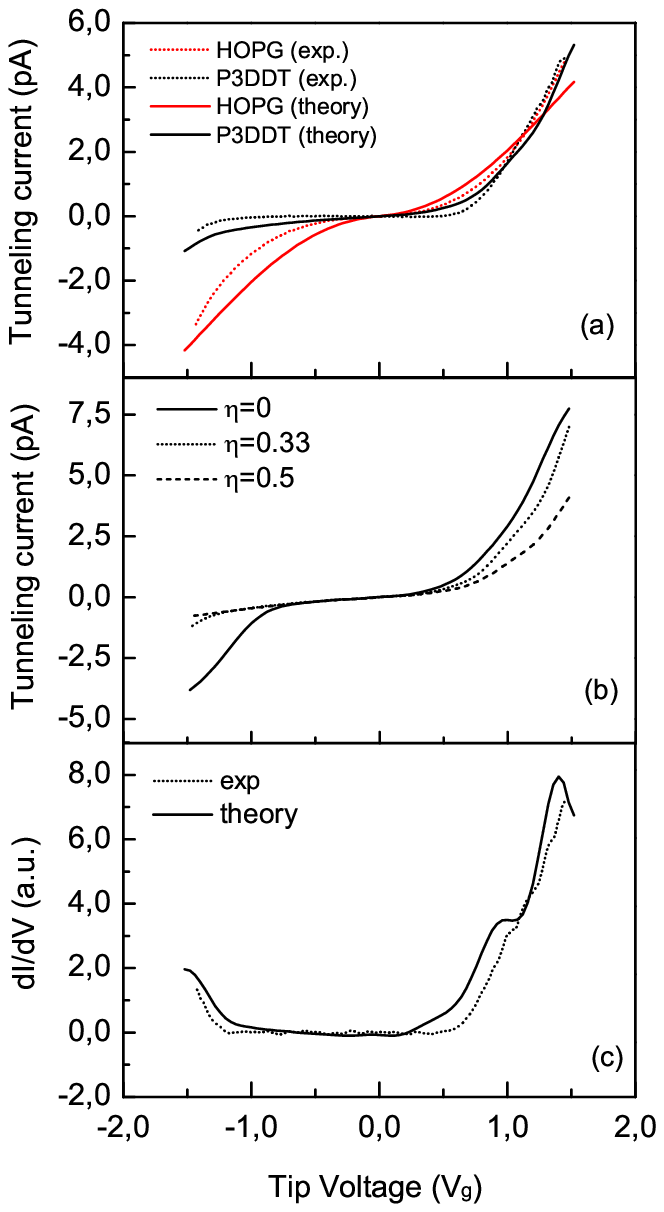}
\caption{(Color Online) (a) Calculated (plain lines) and experimental (dotted lines) $I(V_g)$ curves for the bare substrate (red) and the polymer chain (black). The calculated curve for the octadecathienyl molecule is presented for $\eta=0.33$ (see text). (b) Calculated $I(V_g)$ curves of the polymer chain for different values of $\eta$. (c) Theoretical and experimental conductance $dI/dV_g$ spectra obtained for P3DDT. The presented calculated spectra has been obtained for $\eta=0.33$.}
\label{curves}
\end{figure}

As a first application of our model, $I(V_g)$ curve has been calculated for the bare graphene surface. In these calculations, the STM tip is positioned at a height giving a value of the tunneling current of the order of the experimental one (some pA, for a tip height of about 8 {\AA} above the surface in the simulations). The comparison with experimental measurements is shown on Fig.\ref{curves}a with a good agreement, both curves being consistent with the semi-metallic nature of graphite. A completely different behavior is found for the polymer chain for which $I(V_g)$ curves and conductance spectra exhibit an extended plateau (or conductance gap) with zero current. Calculations have been performed for severals values of the $\eta$ parameter between $0.5$ and $0$. We remind that this parameter is used to describe how the potential falls through the junction due to the coupling of the molecule with the substrate and the STM tip. The value $\eta=0.5$ indicates that the molecule is equally interacting with the two electrodes, whereas the case $\eta=0$ reflects a predominant coupling with the surface. Values between $0.5\leq \eta \leq 1$ should be excluded since a stronger interaction with the STM tip than with the substrate seems unreasonable. In Fig.\ref{curves}b, the calculated $I(V_g)$ curves for three different values of $\eta$ ($0$, $0.33$ and $0.5$) are shown (since the position of the tip is kept constant for all the calculations, the three presented curves can directly be compared to each other). The shape of these curves is very similar and a zero conductance plateau is obtained for each cases. However, the width of this plateau is directly related to the value of $\eta$, which suggests that the abrupt increase of the tunneling current is due to the contribution of the molecular levels. Indeed, the shift of these levels with the applied bias is related to the potential's drop in the junction (Eq.\ref{pot_drop}). Hence, for a high enough negative bias, electrons will be injected from the STM tip to the surface via the $\tilde{EA}$ (Fig.\ref{elec_levels}b) and first unoccupied levels, as we stand in the situation where we have a quasi continuum of states. Similarly, for a sufficiently strong positive bias, electrons will be injected from the surface to the tip via the $\tilde{IP}$ and last occupied levels of the molecule (Fig.\ref{elec_levels}c); the values of these voltages depending on the value of the $\eta$ parameter.  

Another important feature of these $I(V_g)$ curves is their asymmetry as a much higher current is obtained for positive biases than for negative ones. Additionally, the voltage for which a tunneling current suddenly increases is not the same at positive and negative polarity. Differently to the work of Terada et \textit{al.} on P3HT on silicon\cite{Terada05} where a valence band-based conduction takes place whatever the bias voltage polarity, we have demonstrated that the conductance gap for P3DDT on HOPG is directly related to the bandgap of the polymer, as also evidenced for polydiacetylene on the same substrate\cite{Akai03}. The asymmetry of the curves is then explained by the difference of voltages to be applied to inject electrons via the occupied or unoccupied molecular states (Fig.\ref{elec_levels}c).\\

The comparison between the calculated spectra and experimental one is given on Fig.\ref{curves}c. A very good agreement has been obtained for $\eta=0.33$ which is consistent with a stronger interaction of the molecule with the substrate due to the fact that, in the low-current STM mode, the tip-sample distance is relatively large. On the other hand, a value of $\eta$ different from $0$ is also expected since the bonding of the molecule with the substrate is weak which supports the absence of charge transfer effects predicted by our \textit{ab initio} calculations. As a matter of fact, within the band gap (Fig.\ref{curves}c), the density of states takes no finite value, leading to a conductance gap of \textit{ca.} 1.7 eV from experimental curves. In other words, there are no charge transfer effects from the substrate to the polymer for the creation of polaronic states inside the semi-conducting gap.\\
Finally, one can mention that, for quasi one dimensional $\pi$-conjugated systems, $\pi$-band edge singularities might be expected. Such features have been reported from STS spectra on PDA\cite{Akai03}, but the electronic structure of P3DDT, as revealed by experimental and simulated curves, appears more complex. On measured spectra, bumps are observed on the falling edge of the last occupied molecular state instead of a well define single peak. dI/dV calculated at $T=0$ K shows, for its part, several peaks at positive voltages, in good agreement with the measured bumps if one takes into account thermal broadening effects.

\section{Conclusion}

In conclusion, we have presented a generic method that combines first-principles and semi-empirical approaches for the simulation of STM experiments in the case of molecules weakly bonded to surfaces. A tight binding scheme has been developed for the calculation of tunneling current, whereas ab initio calculations have been used to extract information about the structural and electronic properties of the system. Although particularly adapted to long periodic molecules, such as polymer chains, it would be possible, in principle, to apply the same methodology to smaller and non periodic ones. We have applied these methods to the study of defectless poly(3-dodecylthiophene) chains adsorbed on graphite, and showed that STS spectra can directly be related to the electronic structure of the polymer itself. Moreover, the comparison with experimental data has presented a very good agreement and supports the validity of our theoretical approach. Furthermore, the wide number of experimental results published by the scientific community on $\pi$-conjugated systems adsorbed on HOPG, highlights the importance of this theoretical work than can easily be applied to this great variety of systems.\\
The authors would like to thank S. Roche and C. Delerue for helpful discussions. M. Dubois acknowledges financial support from CEA/DRT under grant ACAV, and the UPV/EHU for supporting his stay in San Sebastian. A. Rubio is supported by the Nanoquanta Network of Excellence (NMP4-CT-2004-500198), Spanish McyT and the Humboldt Foundation under the 2005 Bessel research award. This work has also been supported by the Micro and Nanotechnology Program of French Ministry of Research under the grant "RTB : Post CMOS moléculaire 200mm". Dr. R. Baptist, head of this research program at CEA-Grenoble is acknowledged for his support.

\bibliography{biblio}

\end{document}